%% file: main.tex
\documentclass[times]{kristall}	

\usepackage{epsf}

\usepackage[round]{natbib}              

\begin{document}
\maketitle		

\begin{abstract}	
\Abstract		
\end{abstract}		

\input contrib				

\end{document}

%% file: contrib.tex
%
%

\def\St{\large\strut}
\def\dc{C$^\circ$}
\def\tc{T$_{c}$}
\def\oh{$\frac{1}{2}$}
\def\th{$\frac{3}{2}$}


%
\section*{Introduction}
\begin{figure}[bt!]
  \epsfxsize=\linewidth
  \epsfbox{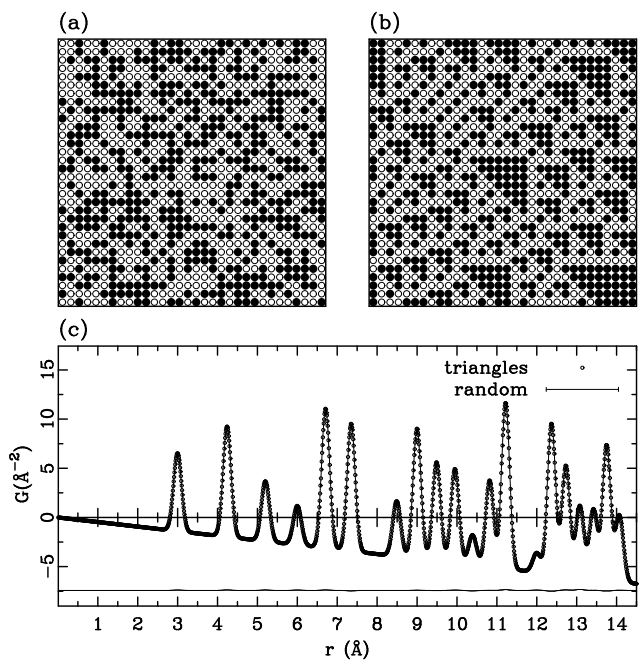}
  \caption[]{(a) Structure with random defect distribution, (b) structure
             showing three-body correlations, but all two-body correlations
             are zero and (c) the resulting PDF of both structures given. The
             difference is magnified by a factor of 10 and plotted below
             the data sets.
            }
  \label{fig;simtri}
\end{figure}

The determination of crystal structures is an important part of
chemistry, physics and of course crystallography. Conventional
structure determination is based on the analysis of the intensities
and positions of Bragg reflections which only allows the determination
of the long range \emph{average} structure of the crystal. For powder
diffraction data this is now routinely done using the Rietveld
\citep{rietv;jac69} method.  It should be kept in mind that the
analysis of Bragg scattering assumes a perfect long range periodicity
of the crystal. However, many materials are quite disordered and even
more important the key to a deeper understanding of their properties
is the study of deviations from the \emph{average} structure or the
study of the \emph{local} atomic arrangements \citep{bi96}. Deviations
from the average structure result in the occurance of diffuse
scattering which contains information about two-body interactions. It
is interesting to remember that the information that can be extracted
from diffuse scattering is indeed limited to two-body correlations. In
an review article about diffuse scattering, \citet{webu94} show two
chemically disordered structures giving rise to the same diffuse
scattering pattern. Two similar structures are shown in Figure
\ref{fig;simtri} and one can easily see the different arrangement of
the two atoms types shown as filled and empty circles. However, for
both structure all two body correlations are zero, the only difference
is a non-zero three-body term for one of the structures, giving rise
to the triangular arrangement seen in Figure
\ref{fig;simtri}. In principle those two structures can not be
distinguished using diffuse scattering or PDF measurements. The
corresponding PDFs for the two structures are shown in Figure
\ref{fig;simtri}c. However, in many cases three-body correlations
might impose constrains on local displacements which then change the
diffraction pattern or PDF. A detailed discussion of this topic can be
found in \citet{we86}.

There are various ways to extract information about the defect
structure from single crystal diffuse scattering.  An overview of
traditional approaches and the analysis of diffuse scattering via
computer simulations can be found in \citep{webu94}, further general
information about disorder diffuse scattering can be found in numerous
review articles \citep{we85,ja87,jafr93,fr95,fr97,webu95}. However,
the data collection, interpretation and analysis of diffuse scattering
is a time consuming and in general difficult task. Information about
the \emph{local} environment of atoms in a given structure can be
extracted using extended X-ray fine absorption structure (XAFS)
analysis and related techniques. A recent review of those methods was
given by \citet{yacoby;currop99}.  The method to reveal the local
structure of crystals used here is the analysis of the PDF.  This
approach is long known in the field of studying short range order in
liquids and glasses but has recently been applied to crystalline
materials \citep{egami;lsfd98,toeg92,bieg93}.  The PDF is obtained
from the powder diffraction data via a sine Fourier transform of the
normalized total scattering intensity $S(Q)$:
\begin{eqnarray}
  G_{e}(r) & = & 4 \pi r [ \rho(r) - \rho_{0} ] \nonumber \\
           & = &  \frac{2}{\pi} \int_{0}^{\infty} Q[S(Q)- 1]\sin(Qr)dQ,
  \label{eq_four}
\end{eqnarray}
where $\rho(r)$ is the microscopic pair density, $\rho_{0}$ is the
average number density and $Q$ is the magnitude of the scattering
vector. For elastic scattering $Q=4\pi \sin (\theta) / \lambda$ with
$2\theta$ being the scattering angle and $\lambda$ the wavelength of
the radiation used. Since the total scattering $S(Q)$ contains Bragg
and diffuse scattering, the information about \emph{local}
arrangements is preserved. Details about the determination of an
experimental PDF can be found e.g. in \citet{egami;lsfd98,warren} and
are not discussed here.

In many systems one is faced with chemical short range order and
subsequent displacements of neighboring atoms. In previous work, the
capability of the RMC technique to refine single crystal diffuse
scattering of systems showing similar disorder was investigated
\citep{prwe97a,prwe97b}.  However, obtaining a PDF from a powder
diffraction experiment is often simpler than carrying out a single
crystal diffuse scattering measurement and allows one to e.g. collect
data for many temperature points around a phase transition or explore
a family of compounds more systematically.  The aim of this paper is
to investigate the capability and possible limitations of reverse
Monte Carlo (RMC) refinement of PDF data to obtain information about
chemical ordering and subsequent displacements.
%
%
\section*{Simulation techniques}
In this section some details about the simulation techniques in general
are given. In the first part the calculation of a PDF from a structural
model is discussed, the following two parts describe the Monte Carlo
(MC) and reverse Monte Carlo (RMC) simulation technique. Although this
paper concentrates on the RMC refinements of various simulated structures
showing occupational and displacive disorder, these defect structures 
themselves were created using the MC method. 
%
%
\subsection*{Calculating the PDF\label{sec;pdf}}
The PDF can be understood as a bond-length distribution between all
pairs of atoms $i$ and $j$ within the crystal (up to a maximum
distance), however each contribution has a weight corresponding to the
scattering power of the two atoms involved. The PDF can be calculated
from a structural model using the relation
\begin{equation}
  G_{c}(r) = \frac{1}{r} \sum_{i}\sum_{j} \left [
             \frac{b_{i}b_{j}}{\langle b \rangle ^{2}}
             \delta (r - r_{ij}) \right ]   - 4 \pi r \rho_{0},
  \label{eq;igr}
\end{equation}
where the sum goes over all pairs of atoms $i$ and $j$ within the
model crystal separated by $r_{ij}$. The scattering power of atom $i$
is $b_{i}$ and $\langle b \rangle$ is the average scattering power of
the sample. In the case of neutron scattering $b_{i}$ is simply the
scattering length, in the case of x-rays it is the atomic form factor
evaluated at a user defined value of $Q$.  The default value is $Q=0$
in which case $b_{i}$ is simply the number of electrons of atom
$i$.\par

Generally there are two different ways to account for displacements
(either thermal or static) from the average position. First one can
use a large enough model containing the desired displacements and
perform an ensemble average. This can be achieved e.g. by displacing
atoms in the model according to a given Debye-Waller
 factor. Alternatively one can convolute each delta-function in
(\ref{eq;igr}) with a Gaussian accounting for the displacements. The
width $\sigma_{ij}$ of the Gaussian is given by the anisotropic
thermal factors $U_{lm} = \langle u_{l}u_{m} \rangle$ of atoms $i$ and
$j$. 

The study of a measured PDF ranges from a simple peak width analysis
revealing information about correlated motion \citep{jeprmjbi99} to
the full profile refinement of the PDF based on a structural model
either using the Reverse Monte Carlo technique
\citep{toeg92,nimckeha95} or least square regression
\citep{billi;b;lsfd98}. The later can be carried out using the program
\emph {PDFFIT} \citep{prbi99}. The reader might refer to this paper
for details on the calculation of PDFs and experimental factors like
terminating the data at a given value of $Q_{\mbox{max}}$ determined
by the experiments. We will exclude the discussion of experimental
influences on the PDF from this paper, since its scope is a principle
investigation of the capability of RMC refinements of the PDF to
obtain information about chemical ordering and possibly accompanying
displacements. All simulations and refinements presented in this paper
were carried out using the program \emph{DISCUS} \citep{prne97}.
%
%
\subsection*{Monte Carlo (MC)\label{sec;mc}}
The total energy of a model crystal is expressed as a function of random
variables such as site occupancies or displacements from the average
structure. The algorithm is based on the original Metropolis Monte
Carlo technique \citep{merorotete53}. A site within the crystal
is chosen at random and the associated variables are altered by some
random amount. The energy difference $\Delta E$ of the configuration before
and after the change is computed. The new configuration is accepted if
the transition probability $P$ given by equation \ref{eq;mc} is greater than a
random number h, chosen uniformly in the range [0,1]. 
\begin{equation}
    P = \frac{\exp(-\Delta E / kT)}{1 + \exp(-\Delta E / kT)}
\label{eq;mc}
\end{equation}
In \ref{eq;mc} $T$ is the temperature and $k$ Boltzmann's constant. It
should be noted that the value of the temperature $T$ controls the
proportion of accepted modifications which lead to a larger total
energy. The process is repeated until the system reaches its
equilibrium. In this paper we refer to a single MC (or RMC) step as
'move', whereas the number of moves necessary to visit every crystal
site once on average will be called one 'cycle'.

Next we briefly describe the actual energy expressions used to
introduce occupational and displacement disorder using the MC
algorithm. All simulations start from a given undistorted average
crystal structure. Chemical ordering is described using binary random
variables $\sigma_{ij}$, with $\sigma_{ij}=+1$ to represent a Cu atom
and $\sigma_{ij}=-1$ for a Au atom. The energy of interaction between
sites used in the MC vacancy ordering scheme is of the form of
\begin{equation}
  E_{\mbox{occ}} = \sum_{i,j} \sigma_{ij}[ H +
                   J_{1}\sigma_{\langle 100 \rangle} +
                   J_{2}\sigma_{\langle 110 \rangle} + ..]
\label{eq;occ}
\end{equation}
where $\sigma_{\langle 100 \rangle}$ is the sum of all four
nearest-neighbor variables, $\sigma_{\langle 110 \rangle}$ the sum of
all four next-nearest-neighbor variables and so on. The sum in
equation \ref{eq;occ} is over all crystal sites $i$ and $j$. The
interaction parameters $H$, $J_{1}$ and $J_{2}$ are initially unknown
and a feedback mechanism is used to achieve the desired ordering of
the vacancies\citep[see][]{webu94}. The displacements were modeled in
a similar way using a Hamiltonian, where the atoms move in harmonic
potentials (Hooke's law).

\begin{equation}
  E_{\mbox{disp}} = \sum_{i,j} k(d_{ij} - \tau_{ij}d_{0})^{2}
\label{eq;disp}
\end{equation}

The sum is over all atoms $i$ and all nearest-neighbor atoms
$j$. Only nearest-neighbor terms are used in this modeling. The
atom-atom distance is given by $d_{ij}$, the average distance is
$d_{0}$, $\tau_{ij}$ is the displacement factor and $k$ a force
constant. 
%
%
\subsection*{Reverse Monte Carlo (RMC)\label{sec;rmc}}
The RMC method was originally developed by \citet{mcpu88}. The
principle is the same MC type algorithm described in the last
section. However, rather than minimizing the total energy, the
difference between observed and calculated data (in our case PDFs) is
minimized.  The algorithm works as follows: First, the PDF is
calculated from the chosen crystal starting configuration and a
goodness-of-fit parameter $\chi^{2}$ is computed.

\begin{equation}
	\chi^{2} = \sum_{i=1}^{N} 
	           \frac{ ( G_{e}(r_{i}) - G_{c}(r_{i})) ^{2}}
	                {\sigma^{2}}
	\label{eq;rmc}
\end{equation}

\begin{figure}[tb!]
  \epsfxsize=\linewidth
  \epsfbox{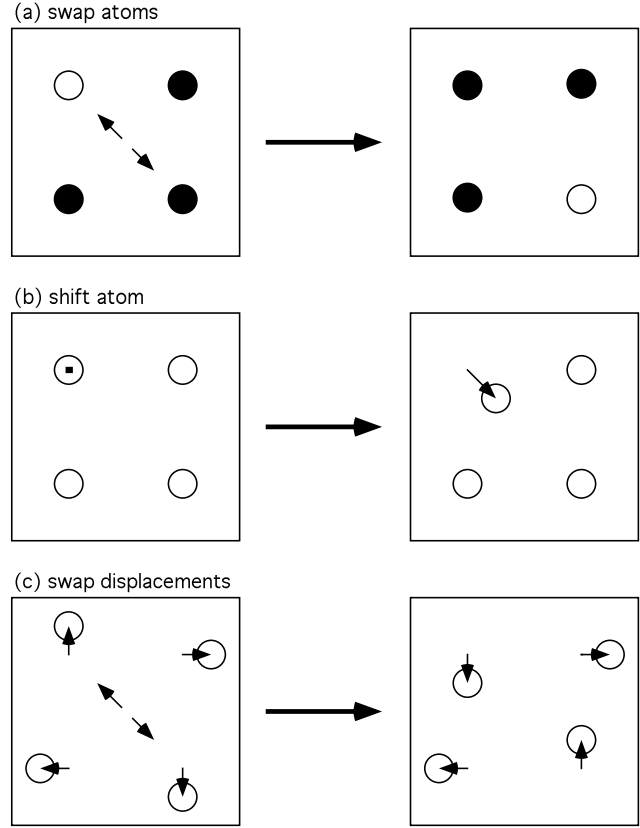}
  \caption[]{Illustration of the MC/RMC modes: (a) swap atoms,
             (b) shift atom and (c) swap displacements from average
             position.}
  \label{fig;rmcmodes}
\end{figure}

The sum is over all measured data points $r_{i}$, $G_{e}$ stands for
the experimental and $G_{c}$ for the calculated PDF.  The RMC
simulation proceeds with the selection of a random site within the
crystal.  The system variables associated with this site, such as
occupancy or displacement, are changed by a random amount, and then
the model PDF and the goodness-of-fit parameter $\chi^{2}$ are
recalculated.  The change $\Delta\chi^{2}$ of the goodness-of-fit
$\chi^{2}$ before and after the generated move is computed.  Every
move which improves the fit ($\Delta\chi^{2} < 0$) is accepted.  'Bad'
moves worsening the agreement between the observed and calculated PDF
are accepted with a probability of $P=\exp(-\Delta\chi^{2}/2)$.  As
the value of $\Delta \chi^{2}$ is proportional to $1 / \sigma^{2}$,
the value of $\sigma$ has an influence on the amount of 'bad' moves
which will be accepted.  Obviously there are two extremes: For very
large values of $\sigma$, the experimental data are ignored ($\chi^{2}
\approx 0$) and with very small values of $\sigma$ the fit ends up in
the local minimum closest to the starting point, because there is a
negligible probability for 'bad' moves.  The parameter $\sigma$ acts
like the temperature $T$ in 'normal' MC simulations.  The RMC process
is repeated until $\chi^{2}$ converges to its minimum.  The result of
a successful RMC refinement is {\it one} real space structure which is
consistent with the observed PDF. In order to exclude chemically
implausible resulting structures additional constrains, e.g. minimal
allowed distances between atoms, may be introduced.

The program used for the RMC refinements presented in this paper is
{\it DISCUS} \citep{prne97}. The program is capable of modeling
occupational as well as displacement disorder and so far we have
called each crystal modification simply RMC move.  In practice we use
a mode ('switch-atoms') of simulation in which occupational disorder
is modeled by swapping two different randomly selected atoms
(Fig. \ref{fig;rmcmodes}a).  This procedure forces the relative
abundances of the different atoms within the crystal to be constant.
It should be noted, that vacancies are treated as an additional atom
type within the program {\it DISCUS}.  The introduction of
displacement disorder can be realized in two different ways.  In the first
method a randomly selected atom is displaced by a random Gaussian
distributed amount ('shift', Fig. \ref{fig;rmcmodes}b).  Alternatively
the displacement variables associated with two different randomly
selected atoms are interchanged ('switch-displacements',
Fig. \ref{fig;rmcmodes}c).  The latter method has the advantage that
the overall mean-square displacement averages for each atom site can
be introduced into the starting model and these will remain constant
throughout the simulation.
%
%
\section*{Occupational disorder: Cu$_{3}$Au}
\begin{figure}[tb!]
  \epsfxsize=\linewidth
  \epsfbox{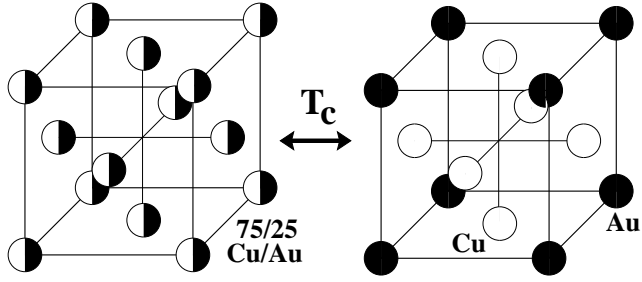}
  \caption[]{Schematic drawing of the structure of Cu$_{3}$Au above and
             below the ordering phase transition at \tc=394\dc.
            }
  \label{fig;cuaustru}
\end{figure}

In this section we will extract chemical ordering information from a
simulated PDF, which we will refer to as 'data'.
This allows one to test the capability of the PDF method to extract such
information for {\it crystalline} samples. The basis for these
simulations is the system Cu$_{3}$Au. The chemical ordering in these
systems has been studies for quite some time and features as text book
example for chemical ordering e.g. in the highly recommended book by
\citet{cowley}. The structure of Cu$_{3}$Au is cubic, space group
$Fm\overline{3}m$ with a lattice parameter of $a=3.734$\AA\ at room
temperature. The system undergoes a order-disorder phase transition at
\tc=394\dc, schematically shown in Figure \ref{fig;cuaustru}. Above
\tc\ the two atom types are randomly distributed, below \tc\ copper
atoms occupy the corners of the unit cell and gold occupies the face
centers.
%
%
\subsection*{Simulations}
Chemical ordering ordering is most conveniently described using the
correlation coefficient $c_{ij}$ which is defined as:
\begin{equation}
  c_{ij} = \frac {P_{ij} - \theta^{2}} { \theta (1 - \theta)}
  \label{eq;corr}
\end{equation} 
$P_{ij}$ is the joint probability that both sites $i$ and $j$ are
occupied by the same atom type and $\theta$ is its overall occupancy.
Negative values of $c_{ij}$ correspond to situations where the two
sites $i$ and $j$ tend to be occupied by {\it different} atom types
while positive values indicate that sites $i$ and $j$ tend to be
occupied by the {\it same} atom type.  A correlation value of zero
describes a random distribution.  The maximum negative value of
$c_{ij}$ for a given concentration $\theta$ is $-\theta/(1-\theta)$
($P_{ij}=0$), the maximum positive value is +1 ($P_{ij}=\theta$).
\begin{figure}[tb!]
  \epsfxsize=\linewidth
  \epsfbox{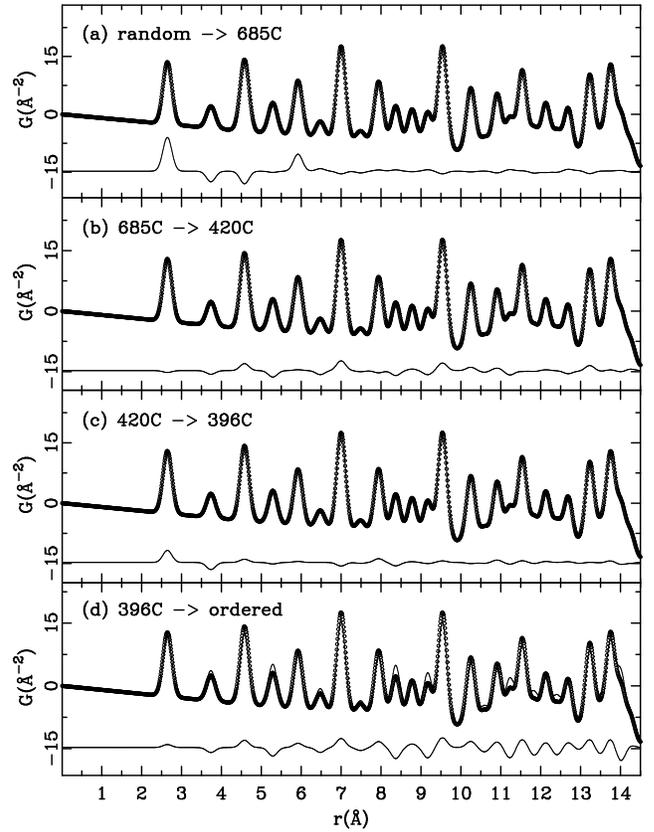}
  \caption[]{Change of the calculated PDFs for Cu$_{3}$Au as a function
             of temperature. The PDFs were generated from the correlation
             values listed in Table \ref{tab;cuau} using MC simulations.
             The difference is plotted below the data sets in each panel.
             Note that the differences for panels (a) to (c) are enlarged
             by a factor of 15, whereas the difference shown in panel (d)
             is not enlarged.
            }
  \label{fig;cuaupdf}
\end{figure}
\begin{table*}[!bt]
\caption[]{Cu$_{3}$Au refinement results.}
\small
\begin{tabular*}{\linewidth}{@{\extracolsep{\fill}} @{}crrrrrrrrrrr}
\hline
     &  \multicolumn{2}{c}{\textbf{ordered}}  &
        \multicolumn{3}{c}{\textbf{T=396\dc}} &
        \multicolumn{3}{c}{\textbf{T=420\dc}} &
        \multicolumn{3}{c}{\textbf{T=685\dc}} \St\\
     Neighbor & Model & RMC &
              Input & MC & RMC &
              Input & MC & RMC &
              Input & MC & RMC \St\\
\hline
\oh\oh 0&-0.33&-0.30&-0.18&-0.17&-0.17&-0.13&-0.13&-0.13&-0.13&-0.13&-0.13\St\\
100     & 1.00& 0.91& 0.21& 0.23& 0.23& 0.15& 0.16& 0.16& 0.11& 0.12& 0.12\St\\
1\oh\oh &-0.33&-0.30& 0.01& 0.01& 0.01& 0.02& 0.02& 0.02& 0.03& 0.04& 0.04\St\\
110     & 1.00& 0.90& 0.06& 0.07& 0.07& 0.05& 0.06& 0.06& 0.02& 0.01& 0.00\St\\
\th\oh 0&-0.33&-0.29&-0.08&-0.08&-0.08&-0.08&-0.07&-0.07&-0.07&-0.07&-0.07\St\\
111     & 1.00& 0.90& 0.02& 0.03& 0.02& 0.01& 0.03& 0.03&-0.01&-0.03&-0.04\St\\
\th 1\oh&-0.33&-0.30&-0.01&-0.01&-0.01&-0.01&-0.02&-0.02& 0.00& 0.01& 0.01\St\\
200     & 1.00& 0.89& 0.07& 0.08& 0.08& 0.07& 0.06& 0.07& 0.03& 0.04& 0.05\St\\
\th\th 0&-0.33&-0.29&-0.03&-0.01&-0.01&-0.02& 0.00& 0.00& 0.00&-0.01&-0.01\St\\
2\oh\oh &-0.33&-0.29& 0.03&-0.01&-0.01& 0.02& 0.00& 0.00& 0.01&-0.01&-0.01\St\\
210     & 1.00& 0.89& 0.03& 0.05& 0.04& 0.02& 0.03& 0.03& 0.00& 0.00& 0.00\St\\
\hline
\end{tabular*}
\label{tab;cuau}
\end{table*}
Using MC simulations as described earlier, a total of five
disordered structure were created corresponding to temperatures of:
below \tc\ (ordered), T=396\dc, 420\dc, 685\dc\ and random distribution
of Cu and Au. The simulated crystal structures extended over 10x10x10
unit cells, containing a total of 4000 atoms. The concentration of
gold is as expected $\theta=0.25$. The target correlation values were
taken from \citet{chcoco79} and all eleven neighbor values were used
in the simulation. The MC simulations were carried out for 200 cycles,
i.e. every atom was visited on average 200 times. The target values
and achieved values after the MC simulations are shown in Table
\ref{tab;cuau} in columns 'input' and 'MC', respectively. Note that
the ordered structure of Cu$_{3}$Au was created directly from the
structural information. The PDFs calculated from these structures are
shown in Figure \ref{fig;cuaupdf}. The calculation was done for x-ray
scattering. In order to observe the change in the PDF more clearly,
PDFs of two subsequent temperatures and the resulting difference are
plotted as separate panels in Figure \ref{fig;cuaupdf}. At high
temperature the Cu and Au atoms are randomly distributed, which
corresponds to correlation values $c_{ij}=0.0$ for all neighbors. At
T=685\dc\ we are still well above the phase transition at
\tc=394\dc\ and inspection of Table \ref{tab;cuau} shows a significant 
negative correlation for the nearest neighbor and a positive
correlation for the next-nearest neighbor. At this temperature the
system starts having preferred Cu-Au nearest neighbors, however the
ordering does not yet extend to larger neighbor distances. The top
panel in Figure \ref{fig;cuaupdf} shows the PDFs for the random
arrangement and the situation at $T=685$\dc. The main differences that
can be observed for the first few neighbors showing correlations
significantly different from zero. As the temperature is lowered
towards the phase transition, the differences of the PDFs between
temperature points become smaller, however, as correlations of more
distant neighbors deviate from zero, changes in the PDF at higher
values of $r$ can be seen. Note that the difference curves of the top
three panels in Figure \ref{fig;cuaupdf} are enlarged by a factor of
15. The bottom panel of Figure \ref{fig;cuaupdf} show the difference
between the PDF just above the phase transition and the ordered low
temperature state. Here differences are clearly visible and this time
the difference curve is not enlarged. It can be clearly observed that
the main change of the PDF is not at higher values of $r$, since in
the ordered state, non-zero correlations extend over the complete
crystal.

The aim of these simulation was to explore the capability of RMC
simulations of these calculated PDFs to recover the correct chemical
correlation values at least for this idealized situation not
considering experimental factors. The RMC simulation method was
described previously. The parameter $\sigma$ given in equation
\ref{eq;rmc} was decreased linearly during the RMC refinement. This
had the effect that initially the ratio of accepted 'good' to 'bad'
moves was about 2:1 which then decreased until no 'bad' moves were
accepted in the last cycles. The refinements were carried out in
'swchem' mode which kept the relative abundance of Cu and Au constant
at its initial values. The refinement was stopped after 15 RMC cycles
when practically no RMC moves were accepted.
%
%
\subsection*{Results}
\begin{figure}[tb!]
  \epsfxsize=\linewidth
  \epsfbox{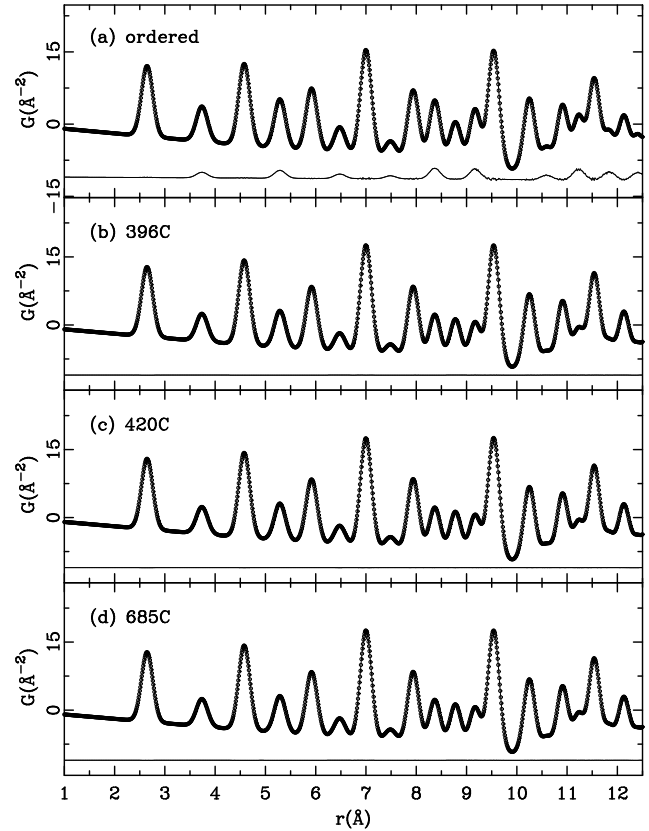}
  \caption[]{Result of RMC refinements of Cu$_{3}$Au. The differences
             are magnified by a factor of 5 and plotted below the data 
             in each panel. For details see text.
            }
  \label{fig;cuaurmc}
\end{figure}
The results of the PDF refinements for the ordered structure and the
three temperature points above \tc\ are shown in Figure
\ref{fig;cuaurmc}. The corresponding correlation values are given in
Table \ref{tab;cuau} in the column marked 'RMC'. Inspection of Figure
\ref{fig;cuaurmc} reveals a nearly prefect agreement between the
simulated 'data' and the RMC result for all data sets except for some
small differences in case of the ordered low-temperature structure. It
should be noted, that the differences shown in Figure
\ref{fig;cuaurmc} are enlarged by a factor of 5. When comparing the
resulting correlation values, one finds a very good agreement between
the initial correlations (column 'MC') and the refinement results
(column 'RMC'). The most noticeable differences again are for the
ordered structure. It is, however, not surprising that the RMC
refinement is not able to produce the ordered structure exactly
without some residual deviations. The results clearly show that
chemical ordering information from crystalline materials can be
obtained using RMC refinements of PDF data. However, one should keep
in mind, that these simulations were carried out to investigate the
principle possibilities of this technique, excluding problems
introduced by noise and systematic errors one would encounter when
using real experimental data.
%
%
\section*{Occupational and displacive disorder}
The next step is to add distortions to a defect structure showing
chemical short range order. In this example, the simulated test
structure had no 'real' analog. For convenience, we used the same
concentration of Cu and Au atoms as in Cu$_{3}$Au.
%
%
\subsection*{Simulations}
\begin{figure}[tb!]
  \epsfxsize=\linewidth
  \epsfbox{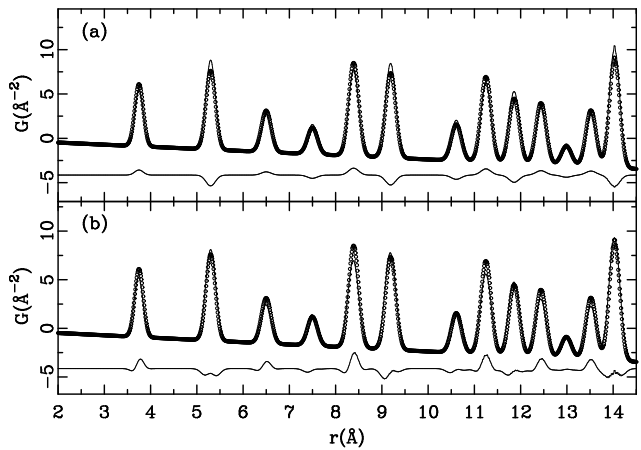}
  \caption[]{PDFs of simulated structures: (a) occupational disorder
             only and (b) occupational and displacive disorder. In both
             panels the PDF of the undistorted structure is shown for 
             reference. The differences are plotted below the data 
             in each panel. For details see text.
            }
  \label{fig;simpdf}
\end{figure}
The disordered structures were created in two steps: First the desired
chemical ordering was introduced the same way as described in the
previous section. As a second step, the atoms were displaced using MC
simulations and a Hooke's law potential as described above. The
starting structure used is cubic primitive with a lattice parameter of
$a=3.75$\AA. The model crystal extends over 32x32x32 unit cells giving
a total of 32768 atoms. The site at (0,0,0) is randomly occupied with
75\% copper atoms and 25\% gold atoms.  The chemical arrangement is
characterized by a negative nearest neighbor correlation of
$c_{100}=-0.3$ and a positive second nearest neighbor correlation of
$c_{110}=0.5$. In other words nearest neighbor Cu-Cu and Au-Au pairs
are avoided whereas second nearest neighbor pairs of the same atom
type are favored compared to a random arrangement. The correlation
coefficients actually achieved by the MC simulation are listed in
Table \ref{tab;disp}. The PDF calculated from the structure containing
only the occupational short-range order is shown in Figure
\ref{fig;simpdf}a) superimposed on the PDF of the random structure. 
The difference between both data sets is plotted as well.

The effect of thermal motion on the PDF is simulated by convoluting
with a Gaussian. In addition the atoms are shifted from their ideal
position by the MC simulation introducing a size-effect type
\citep{buwiwe92} distortion. In this case the resulting nearest
neighbor distances are is $d_{\mbox{cu-au}}=3.73$\AA\ and
$d_{\mbox{cu-cu,au-au}}=3.77$\AA\ (Table \ref{tab;disp}). Note that
the average distance is the lattice parameter $a=3.75$\AA. A total of
200 MC cycles were computed to create this structure. The calculated
PDF is shown in Figure \ref{fig;simpdf} in the bottom panel. The PDF
of the undistorted starting structure is plotted in the same panel and
the difference between both is shown below the data. 
\begin{figure}[tb!]
  \epsfxsize=\linewidth
  \epsfbox{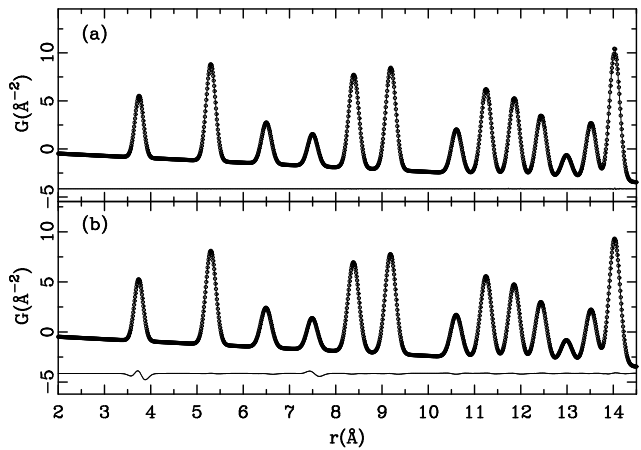}
  \caption[]{Results of RMC refinements: (a) occupational disorder
             only and (b) occupational and displacive disorder. The 
             differences between are magnified by a factor of 5 and 
             are plotted below the data in each panel. For details 
             see text.
            }
  \label{fig;simrmc}
\end{figure}
\begin{table}[!tb]
\caption[]{Results of RMC refinements of disordered structures
           showing occupational as well as displacive disorder.
           The bond lengths are given in \AA.}
\small
\begin{tabular*}{\linewidth}{@{\extracolsep{\fill}} @{}crrr}
\hline
Neighbor               &  Input   &   Run O &    Run D    \St\\
\hline
1 0 0                  &  -0.30   &  -0.29   &  -0.25     \St\\
1 1 0                  &   0.51   &   0.50   &   0.50     \St\\
d$_{\mbox{cu-au}}$     &   3.73   &    -     &   3.74     \St\\
d$_{\mbox{cu-cu,au-au}}$     
                       &   3.77   &    -     &   3.76     \St\\
\hline
\end{tabular*}
\label{tab;disp}
\end{table}

Two different RMC refinements were carried out: Run O using the 'data'
containing chemical ordering only and Run D using the 'data'
containing additional displacements. Run O was carried out as a
reference of the chemical correlations that can be refined when no
displacements are present. The refinement was carried out exactly as
described in the last section. The resulting PDFs are shown in Figure
\ref{fig;simrmc} and the achieved correlation values for run O are
listed in Table\ref{tab;disp}. As for Cu$_{3}$Au, we observe an
excellent agreement between 'data' and the calculation. The
interesting question is now, can one extract the chemical short range
order equally well with the size-effect distortion present, and can one
extract the magnitude of those distortions. In order to model
occupational and displacive disorder simultaneously during the RMC
refinement, we employed the following strategy which was already found
to be successful when refining single crystal diffuse scattering
\citep{prwe97b}. The occupational shifts and displacement shifts were
carried out alternately with no more than 10\% of the crystal sites
being visited before switching between the two modes. As before, the
parameter $\sigma$ was decreased during the RMC refinement in such a
way that initially a significant amount of 'bad' moves was accepted,
however, in the final few cycles, no moves worsening the agreement
between observed and calculated PDFs are accepted. The simulations
were carried out for a total of 15 cycles when the agreement did not
significantly improve any further. The resulting PDFs of run D are
shown in Figure \ref{fig;simrmc} and the achieved correlation values
are listed in Table \ref{tab;disp}.
%
%
\subsection*{Results}
Inspection of Figure \ref{fig;simrmc}a shows a very good agreement
between the RMC result and the 'data' for run O. The chemical
correlations of the resulting structure match the expected ones nearly
perfectly (see Table \ref{tab;disp}) similar to the result found for
the Cu$_{3}$Au simulations presented in the last section. The focus in
this part is on run D where chemical and displacive contributions were
present.  The refined PDF shows a good agreement with the 'data'
(Fig. \ref{fig;simrmc}b), the only noticeable differences are at the
nearest neighbor distance and at twice that distance. This indicates
some residual differences in the exact distribution of distortions
within the crystal. It is interesting to note that the successful
refinement of a system showing displacements required a much larger
model containing in excess of 32000 atoms compared to only 4000 used
for the simulations of Cu$_{3}$Au. In addition to the static
displacements introduced as size-effect, the contribution of thermal
vibrations to the PDF needs to be accounted for. Various test runs
showed that modeling the thermal component via convolution of the PDF
with a Gaussian gives better results compared to randomly displacing
all atoms in the model structure. Inspection of the achieved
correlations and displacements listed in Table \ref{tab;disp} shows an
overall good agreement with the expected values, although the nearest
neighbor correlation does not match as good as for run D compared to
the result of run O. The resulting distortions match the expected
values as well. In summary, the chemical correlations as well as
magnitudes of distortions were successfully extracted from the PDF
using RMC simulations.
%
%
\section*{Conclusions and outlook}
\begin{figure}[tb!]
  \epsfxsize=\linewidth \epsfbox{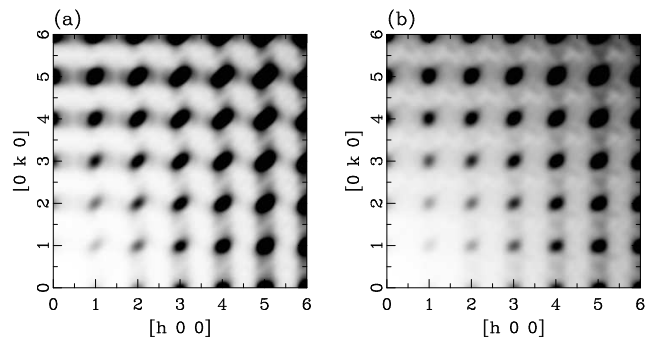} 
  \caption[] {Diffuse scattering from simulated structures: 
  (a) input structure for simulation 'Both', (b) corresponding RMC
  result.  Units are reciprocal lattice parameters. The calculation
  was done using neutron scattering. For details see text.}
  \label{fig;simdiff}
\end{figure}

Two important conclusions can be made from the results presented in
this paper: First the chemical short range order information can be
extracted from the PDF using RMC simulations even in cases where
additional displacive disorder is present. Furthermore, one can also
extract the magnitude of the specific distortions very reliably. For
this simple isotropic disorder, the extracted correlations for the
underlying chemical ordering from the PDF are even slightly closer to
the expected values than in a similar set of RMC refinements of single
crystal diffuse scattering \citep{prwe97b}. It is interesting to
compare the diffuse scattering calculated from the input and resulting
structure of run D. The calculated diffuse scattering for the zeroth
layer of the hk-plane in reciprocal space is shown in Figure
\ref{fig;simdiff}. The patterns show a qualitative agreement between 
'data' and refinement result, which strengthens the results obtained
from refining the PDFs. In case of refining the diffuse scattering
directly one obtains a much better match with the observed data
\citep[see][]{prwe97b}. However, although the diffuse scattering patterns 
shown in Figure \ref{fig;simdiff} are in general similar, closer
inspection reveals an interesting difference: The pattern
corresponding to the input structure has in general sharper more
localized diffuse scattering whereas the pattern calculated from the
RMC result is somewhat broader. This indicated that the defects in the
resulting structure are not correlated over long enough
lengths. Although the PDF contains diffuse scattering information as
well, there is a certain complementarity between single crystal
diffuse scattering and PDF data. We are currently investigating this
relationship on the system In$_{1-x}$Ga$_{x}$As \citep{jeong;unpub00}.
This opens the possibility of a joined refinement of single crystal
diffuse scattering and PDF data obtained from powder data.

The 'data' used for the refinements in this paper were calculated from
given disordered structures, which allowed us to judge the results
since the 'correct' result was known. Additionally those data were
free from experimental errors and noise, which of course will make
refining of PDFs obtained from real experimental data more
challenging. The simulations presented here were carried out using the
total PDF, i.e. all atom pairs are included in the PDF. Using
anomalous scattering or by combining neutron and x-ray data, one can
separate out the different partial PDFs. For a simple binary compound
consisting of atoms A and B, the corresponding partials would contain
only AA, BB or AB (same as BA) atoms. More details can be found in a
recent study of the total and differential PDF of In$_{1-x}$Ga$_{x}$As
by \citet{pejemjprbido99}.  The program \emph{DISCUS} used here can
also be used to calculate or refine any partial PDF of a given
structure. A different definition for partial PDFs was proposed by
\citet{bhth70}. They separated the different terms in a binary case
into a topological term (the usual PDF), the radial concentration
correlation function (RCF) containing information about chemical
fluctuations and a cross-term. Current efforts are to include the
calculation and refinement of the RCF into the \emph{DISCUS} program.

In summary, we have shown in this paper that RMC refinements of the
PDF of disordered crystalline materials can extract chemical
short-range order information for systems showing only occupational
disorder as well as for systems showing occupational and displacive
disorder in combination. The usage of faster and usually simpler
powder diffraction measurements allows one to extract this information
more quickly and cover e.g. many temperature points around a phase
transition. However, for certain problems, one will still need to use
single crystal diffuse scattering to determine the details of the
defect structure or even perform a combined refinement of PDF and
single crystal data. Measurements of the system Cu$_{3}$Au using
synchrotron radiation are planned and will be the first 'real' test
using the PDF method to extract chemical short range order parameters
for this crystalline system.
%
%
\begin{acknowledgment}
The author would like to thank A. Tomic for her contributions during the
inital phase of the project as well as S.J.L. Billinge and V. Petkov 
for many fruitful discussions on the topic of chemical ordering and its
influence on the PDF, which ultimately led to the work presented 
in this paper. I would also like to thank T.R. Welberry for his program
and help to create the structures shown in Figure \ref{fig;simtri}.
This work was supported by DOE through Grant No. DE FG02 97ER45651.
\end{acknowledgment}
%
%